\documentclass[11pt,a4paper]{article}
\pdfoutput=1
\usepackage{jheppub}

\makeatletter
\gdef\@fpheader{\hfill MSUHEP-26-004\par\hfill MS-TP-26-20\par\noindent\makebox[\textwidth][c]{\includegraphics[height=4.0cm]{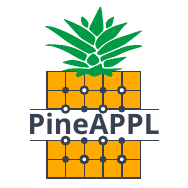}}}
\makeatother

\usepackage[font=small,labelfont=bf,margin=0mm,labelsep=period,tableposition=top]{caption}
\usepackage{latexsym}
\usepackage{graphics}
\usepackage{amssymb}
\usepackage{hyphenat}
\usepackage{amsmath}
\usepackage{relsize}
\usepackage{trimclip}
\usepackage{booktabs} 
\usepackage{tabularx}
\usepackage{multirow}
\usepackage{fixmath}
\usepackage{marginnote}
\usepackage[normalem]{ulem}
\usepackage{siunitx}
\usepackage{cuted}
\usepackage[frozencache=true,cachedir=_minted]{minted}
\usepackage{orcidlink}
\usepackage{pifont}
\usepackage{fontawesome}
\usepackage{mathtools}

\usepackage{tikz}
\usetikzlibrary{calc}
\usepackage{placeins}

\definecolor{lightgray}{gray}{0.9}

\newcolumntype{U}{>{\centering\arraybackslash}X}
\newcommand{\pineappl}{PineAPPL}
\newcommand{\applgrid}{APPLgrid}
\newcommand{\fastnlo}{fastNLO}

\title{\boldmath \pineappl{\sc v1}: fast and flexible theory predictions for present and future colliders}

\author[1]{Tomáš Ježo,~\!\orcidlink{0000-0002-1334-7607}}
\author[2]{Emanuele R. Nocera,~\!\orcidlink{0000-0001-9886-4824}}
\author[3]{Tanjona R. Rabemananjara,~\!\orcidlink{0000-0002-8395-8059}}
\author[]{Christopher Schwan,~\!\orcidlink{0000-0002-8907-5409}}
\author[4]{Tanishq Sharma,~\!\orcidlink{0000-0001-9371-902X}}
\author[1]{Jan Wissmann~\!\orcidlink{0009-0009-1136-0849}}

\affiliation[1]{Institut für Theoretische Physik, Universität Münster, Wilhelm-Klemm-Straße 9, D-48149 Münster, Germany}
\affiliation[2]{Dipartimento di Fisica, Universit\`a degli Studi di Torino, Via Pietro Giuria 1, 10125 Torino, Italy and INFN, Sezione di Torino, Via Pietro Giuria 1, 10125 Torino, Italy}
\affiliation[3]{Universit\'e Paris-Saclay, CNRS, IJCLab, 91405 Orsay, France}
\affiliation[4]{Department of Physics and Astronomy, Michigan State University, East Lansing, MI 48824, USA}

\emailAdd{tomas.jezo@uni-muenster.de}
\emailAdd{emanueleroberto.nocera@unito.it}
\emailAdd{tanjona.rabemananjara@ijclab.in2p3.fr}
\emailAdd{cschwan@posteo.de}
\emailAdd{shar1157@msu.edu}
\emailAdd{jan.wissmann@uni-muenster.de}

\abstract{We present \pineappl\ v1, a library designed to provide accurate and
  flexible interpolation tables of partonic cross sections that can be
  convolved with parton distribution functions (PDFs) and fragmentation
  functions (FFs) for the fast evaluation of high-energy physical observables.
  The core feature of the new release is the support of multiple convolutions
  involving initial- and final-state hadronic particles, with any polarisation,
  that are associated with PDFs and FFs. The library simultaneously supports
  polarised and unpolarised distributions that obey space-like or time-like
  evolution, and is developed for an arbitrary number of them, even if physical
  processes typically only require a few. Control of scale choices when more
  than one scale characterises a scattering process is also possible.
  We describe the technical details of the new representation of interpolation
  coefficients stored in the grid, and we demonstrate the capabilities of the
  library in a few phenomenological cases of interest. Specifically, we compute
  predictions for single-inclusive pion production in unpolarised and polarised
  proton--proton collisions and in semi-inclusive deep-inelastic scattering. We
  show how, in each case, PDF, FF, and scale uncertainties compare to each
  other and highlight the potential of \pineappl\ as an essential ingredient
  for precision physics at current and future colliders.
}

\begin{document}
\maketitle
\flushbottom

\section{Introduction}
\label{sec:introduction}

The efficient computation of cross sections for high-energy hadronic scattering
processes is key to any current and future collider physics programme.
To consolidate our understanding of the Standard Model (SM) and search
for new phenomena in and beyond it, it is essential to compare accurate and
precise theoretical predictions to experimental measurements in the widest
possible kinematic range. The Large Hadron Collider (LHC) operations, including
the high-luminosity phase, have made this demand increasingly compelling.
Because experimental uncertainties are reaching percent or even
sub-percent level, theoretical predictions must match that precision. This
requires, on the one hand, to incorporate higher-order corrections in the
strong and electroweak couplings, and, on the other hand, to push forward
the numerical capabilities of Monte Carlo event generators. Both these
aspects can rapidly render computational costs prohibitive. For instance,
the computation of the $Z$-boson transverse-momentum distribution at
next-to-next-to-next-to-leading order (N$^3$LO) in the strong coupling plus
next-to-next-to-next-to-next-to-leading-logarithmic (N$^4$LL) accuracy
with MCFM~\cite{Campbell:2019dru} requires about $10^5$
CPU hours~\cite{Neumann:2022lft}. Very often, one would like to assess how
theoretical predictions change upon variation of the input distribution
functions\footnote{Throughout this paper, we use the term ``distribution functions'' to collectively refer to parton distribution functions and fragmentation functions.}, for instance when these are compared to the experimental data.
Recomputing predictions every time is obviously very inefficient.

In order to address this issue, several tools have been developed in recent
years, including \applgrid~\cite{Carli:2010rw},
\fastnlo~\cite{Kluge:2006xs, Wobisch:2011ij}, and
\pineappl~\cite{Carrazza:2020gss,christopher_schwan_2025_15635174}.
The idea common to these three libraries 
is to store in some grid format, which varies for each library, the Monte Carlo
phase-space weights corresponding to a fixed-order calculation of a given
partonic cross section. The storage is made memory-efficient by collecting not
the Monte Carlo tuples (each consisting of phase-space variables and weight)
themselves, but rather the
coefficients of a suitably chosen set of polynomial functions used to
interpolate Monte Carlo events on the grid points. The basis of polynomial
functions is chosen in such a way that the original Monte Carlo precision of
the computation is not spoiled by interpolation. This technique allows one to 
greatly reduce the memory footprint of the grid. Any of the three libraries,
when properly interfaced to a Monte Carlo integration program, requires one
to fill the grid only once, independently from the input distribution
functions, the value of the strong coupling, and the
variations of the renormalisation and factorisation scales. The re-evaluation
of hadronic cross sections upon variation of any of these inputs is
then straightforward, as it decouples from the computationally
intensive task of generating Monte Carlo events.

The \pineappl\ library, originally developed by some of us, distinguishes
itself from others in three main aspects. First, it supports the
consistent inclusion of corrections both in the strong and electroweak
couplings, including mixed corrections. Second, it encodes the
dependence on the renormalisation and factorisation scales in the grid,
allowing for their variations a posteriori. Third, it comes with a 
powerful command-line interface (CLI) to perform various
operations on the grids, and bindings to different programming languages. For
instance, inspection of partonic channels and perturbative orders,
variation of scales, computation of scale and input distribution
uncertainties, grid rebinning or merging, export to the
APPLgrid or FastNLO formats, and result visualisation are all easy
operations. \pineappl\ has been interfaced to various Monte Carlo programs such
as {\sc NNLOjet}~\cite{NNLOJET:2025rno,Cruz-Martinez:2025ffa},
{\sc MadGraph5\_aMC@NLO}~\cite{Alwall:2014hca,Frederix:2018nkq},
{\sc MATRIX}~\cite{Grazzini:2017mhc,Devoto:2025cuf}, and
{\sc VRAP}~\cite{Anastasiou:2003ds}.

All of the aforementioned libraries, including \pineappl, however, suffer from
a main limitation: they were
developed targeting the Tevatron and the LHC, therefore they only support
cross sections that involve two unpolarised protons (or a proton and an
antiproton) in the initial state. The \pineappl\ library has been extended
to handle polarised protons in the initial state~\cite{Cruz-Martinez:2025ahf}
and to the simpler case of deep-inelastic-scattering (DIS) by interfacing it to
{\sc yadism}~\cite{Candido:2024rkr,alessandro_candido_2020_3929499,
  Barontini:2024xgu}, including in the polarised case~\cite{Hekhorn:2024tqm},
and for neutrino DIS~\cite{Candido:2023utz,Cruz-Martinez:2023sdv,
  MammenAbraham:2024gun}. There exist, however, a large body of experimental
measurements for semi-inclusive cross sections, in which a hadron in the
final state is detected. These are attracting an increasing interest from
the community because of a significant improvement in experimental precision,
and the launch of dedicated experimental programmes, such as the Electron-Ion
Collider~\cite{Accardi:2012qut,AbdulKhalek:2021gbh}. At the same time, the
accuracy of perturbative computations has progressed steadily:
next-to-next-to-leading order (NNLO) corrections to unpolarised and
polarised semi-inclusive deep-inelastic scattering
(SIDIS)~\cite{Bonino:2024qbh,Bonino:2024wgg,Bonino:2024adk,Bonino:2025tnf,
  Bonino:2025qta,Bonino:2025bqa,Goyal:2023zdi,Goyal:2024tmo,Ahmed:2024owh,
  Goyal:2024emo,Goyal:2025bzf,Goyal:2026ccx},
to single-inclusive hadron production in unpolarised proton--proton
collision~\cite{Czakon:2025yti}, and to hadron
in-jet production in electron-positron annihilation~\cite{Bonino:2026dvr}
have been completed recently. These processes offer unique phenomenological
opportunities. For example: they can be used for precision physics studies or
for new physics searches at colliders, in particular by looking at the
formation of hadrons in jets~\cite{Kaufmann:2015hma}; they provide unique
sensitivity to the flavour structure of the proton that cannot be accessed
through inclusive measurements alone~\cite{Borsa:2017vwy};  they are essential
to access helicity-dependent and transverse-momentum-dependent parton
distributions~\cite{Angeles-Martinez:2015sea}; they are key to disentangling
initial-state nuclear effects from final-state hadronisation dynamics, as they
provide a baseline for studying medium-induced modifications in strongly
interacting matter~\cite{Doradau:2024wli}; and they are used to model hadronic
cascades in cosmic-ray interactions with interstellar matter that affect
predictions for secondary particle fluxes relevant to astroparticle physics
measurement interpretation~\cite{Boglione:2026mzz}.
 
In this paper we extend \pineappl\ to handle cross sections that entail
multiple convolutions among different nonperturbative distributions:
unpolarised and polarised parton distribution functions (PDFs) and unpolarised
and polarised fragmentation functions (FFs). All these objects evolve
differently, respectively according to unpolarised or polarised space-like
and time-like DGLAP equations. We specifically extend the data structure to
accommodate an arbitrary number of convolutions, even if physical observables
typically require only a few, which allows \pineappl\ to scale to any process,
with any type of input distribution, including, {\it e.g.}, multi-parton
scattering and multi-hadron correlations in the final state.
New to \pineappl{\sc v1} is also the possibility to store two independent
kinematic scale variables in the grid, and to choose among a set of functional
forms that combine them. This is useful in processes that are characterised
by more than one scale, and for which the renormalisation and
factorisation scales are typically chosen as an analytic expression of two
hard scales. Renormalisation and factorisation scales, separately in the
PDFs and in the FFs, can be chosen and varied independently. The
phenomenological reach of \pineappl\ therefore becomes more powerful, and
can in particular be used as a tool to aid and guide richer phenomenological
studies not only at the LHC, but also at the now completed Relativistic Heavy
Ion Collider (RHIC) experimental programme and at the forthcoming EIC.
\pineappl\ is publicly available at~\cite{Pineappl:online}. We ensure backward
compatibility with \pineappl{\sc v0}.

This paper is organised as follows. In
Sect.~\ref{sec:extension_to_multiple_convolution}, we review the
general principles underlying \pineappl, describe how the internal data
structure is modified to handle multiple convolutions, illustrate the available
scale choices implemented in \pineappl\ when more than one hard scale
characterises a scattering process, and comment on scale variation prescriptions
in processes where initial- and final-state factorisation scales coexist.
In Sect.~\ref{sec:pheno}, we demonstrate the
new features of \pineappl\ by computing predictions of cross sections
corresponding to the inclusive production of an identified hadron in
proton--proton collisions and in SIDIS. We then comment on possible
phenomenological implications that follow from data-theory comparisons.
In Sect.~\ref{sec:conclusions}, we summarise our results
and discuss future developments. The paper is
complemented by two appendices. In Appendix~\ref{app:MC}, we benchmark
\pineappl\ interpolation accuracy by showing that the grid representation does
not result in any accuracy loss in comparison to the native Monte Carlo results.
In Appendix~\ref{sec:usage-pineappl}, we provide a short illustration
of the installation and usage of the \pineappl\ library. Additional up-to-date
details, including examples of usage of the C, C++, Fortran, and Python
programming languages, can be found in the online
documentation~\cite{Pineappl:online}.

\section{Extension to multiple convolutions}
\label{sec:extension_to_multiple_convolution}

In this section, we discuss the core feature that characterises
\pineappl{\sc v1}, namely the support for multiple convolutions among
different hadronic initial- and final-state distribution functions. We first
recall the general principles that underlie \pineappl, which are common to
{\sc v0} and {\sc v1}. After that, we present how the \pineappl\ data structure is
modified to handle multiple convolutions, by focusing, as an example, on
single-inclusive hadron production in proton--proton collisions. We finally
illustrate the available scale choices implemented in \pineappl{\sc v1}
when more than one hard scale characterises a scattering process,
and review scale variation prescriptions when initial- and final-state
distribution functions coexist.

\subsection{General principles}
\label{subsec:general_principles}

Factorisation theorems allow a sufficiently inclusive class of physical
observables to be determined as a convolution between process-dependent
partonic cross sections, which can be computed as a perturbative expansion in
the interaction couplings, and universal distribution functions, which are
determined from global analyses of experimental measurements
and then delivered through PDF interpolation
libraries~\cite{Buckley:2014ana,Rabemananjara:2025unt}.
The central idea underlying fast-interpolation grid libraries such as
\pineappl\ is to separate the Monte Carlo
computation of partonic weights, which is numerically expensive, from the
subsequent convolution with non-perturbative distributions, which is
instead much faster. The advantage of this method is that, for a specified
choice of theoretical settings, weights can be computed only once, stored in
the format of interpolation grids, and re-used with any input distribution
functions to quickly evaluate physical observables.

In \pineappl{\sc v0}~\cite{Carrazza:2020gss,christopher_schwan_2025_14719930},
this strategy was implemented for processes with two hadrons in the
initial state. For each observable bin $\mathcal{O}$, scale $Q^2$,
perturbative order $k$ ($l$) in the strong (electroweak) coupling $\alpha_s$
($\alpha$), renormalisation scale $\mu_R$ (factorisation scale $\mu_F$)
logarithmic contribution $m$ ($n$), and partonic channel $ab$, the fundamental
object to represent was the weight function
\begin{equation}
  W^{(k,l,m,n)}_{ab}(\mathcal{O}, Q^2, x_1, x_2)\,,
  \label{eq:weight}
\end{equation}
where $x_1$ and $x_2$ are the fractions of the initial-state proton momenta
carried by the partons involved in the hard scattering. For each point $i$ in
the phase space, one could define a 4-tuple
\begin{equation}
  \left\{
  Q^2_i, x_1^i, x_2^i, W^{(k,l,m,n)}_{ab}(\mathcal{O}_i, Q^2_i, x_1^i, x_2^i)
  \right\}_{i=1}^N\,,
\end{equation}
with $N$ the total number of points sampled in the phase space. A convolution
would then be performed by evaluating the parton distributions at the stored
values of $x_1$, $x_2$, and $Q^2$, multiplying them by the corresponding stored
weights, and summing over all $N$ tuples:
\begin{align}
  \frac{d\sigma}{d\mathcal{O}}\left(\mathcal{O}, Q^2,\mu_R^2, \mu_F^2 \right)
  & =
  \sum_{a,b}\int dx_1\, dx_2\,
  f_a\!\left(x_1,\mu_F^2\right)
  f_b\!\left(x_2,\mu_F^2\right)
  \frac{d\hat\sigma_{ab}}{d\mathcal{O}}
  \left(
  \mathcal{O}, Q^2, x_1, x_2, \xi_R^2,\xi_F^2
  \right)
  \nonumber\\
  & \approx
  \sum_i \sum_{a,b}
  f_a\!\left(x_1^i, \xi_F^2 Q^2_i\right)
  f_b\!\left(x_2^i, \xi_F^2 Q^2_i\right)
  \frac{d\hat\sigma_{ab}}{d\mathcal{O}}
  \left(
  \mathcal{O}_i, Q^2_i, x_1^i, x_2^i, \xi_R^2, \xi_F^2
  \right)\,
  \label{eq:pp_incl},
\end{align}
with the partonic cross section
\begin{equation}
  \frac{d\hat\sigma_{ab}}{d\mathcal{O}}
  \left(
  \mathcal{O}_i, Q^2_i, x_1^i, x_2^i, \xi_R^2, \xi_F^2
  \right)
  =
  \sum_{\mathclap{k,l,m,n}}\alpha_s^k\!\left(\xi_R^2 Q^2_i\right)\alpha^l
  \log^m\!\left(\xi_R^2 \right)
  \log^n\!\left(\xi_F^2 \right)
  W^{(k,l,m,n)}_{ab}(\mathcal{O}_i, Q^2_i, x_1^i, x_2^i)\,,
  \label{eq:pp_incl_pxs}
\end{equation}
where $\xi_t^2=\mu_t^2/Q^2$, with $t=R,F$.

This way of proceeding is simple, however it does not lead to an efficient
grid format. The grid size grows linearly with the number
of points in the phase space, {\it i.e.}, with the number of Monte Carlo events,
and the convolution eventually becomes limited by the speed at which a large
number of tuples can be read from disk. For this reason, \pineappl{\sc v0}
used an interpolation grid. Instead of storing all Monte Carlo tuples, the
variables $x_1$, $x_2$, and $Q^2$ were mapped to interpolation variables
defined to resolve efficiently both the small-$x$ and large-$x$ regions. Each
Monte Carlo weight was then distributed over the neighbouring interpolation
nodes using as basis functions Lagrange polynomials chosen as
in~\cite{Carli:2010rw}. The resulting grid therefore stored
interpolation coefficients for the weights, Eq.~\eqref{eq:weight},
rather than the original event list. One may think of such coefficients as
\begin{equation}
  A_{j_{Q^2},j_{x_1},j_{x_2}}^{(k,l,m,n,a,b)}\,
\end{equation}
where $j_{Q^2}$, $j_{x_1}$, $j_{x_2}$ label the interpolation nodes associated with
$Q^2$, $x_1$, $x_2$. Additional reweighting
factors were finally used when filling the grid in order to improve the
interpolation accuracy, and were inverted at convolution time.

This construction is naturally three-dimensional in the interpolation
variables: two dimensions correspond to the incoming momentum fractions,
and one to the event scale. This is sufficient for predictions involving two
incoming PDFs. The extension to multiple convolutions, which is the core change
achieved in \pineappl{\sc v1}, however, requires a more flexible
representation. For example, the inclusive production of an identified hadron
in proton--proton collisions involves two incoming PDFs and one final-state FF,
and therefore depends on $x_1$, $x_2$, $z$, and $Q^2$, where $z$ is the
momentum fraction of the fragmenting parton carried by the produced hadron.\footnote{Here we assume that the initial- and final-state factorisation scales have been identified, up to constant prefactors.}
The three-dimensional tuple structure described above, therefore, does not
generalise efficiently to an arbitrary number of initial- or final-state
hadronic distributions, each with its own momentum fraction and possibly its own
evolution type. A naive dense generalisation of the original grid would be
numerically inefficient, since, if each new convolution variable is represented by an
additional interpolation axis, the number of grid points grows multiplicatively
with the number of dimensions. Moreover, most of the resulting hypercube would
be typically empty, because phase-space constraints populate only a restricted
region of the allowed kinematics. A storage format that explicitly retains all
entries of the dense multidimensional array would therefore waste memory,
especially for processes with several convolutions. For these reasons,
\pineappl{\sc v1} implements a new data structure that overcomes all these
difficulties, which we describe next.

\subsection{Data structure and grid representation}
\label{subsec:data_structure}

\pineappl{\sc v1} supports the computation of physical observables
that involve hadrons, possibly longitudinally polarised, both in the initial
and final states. This means that a different number of PDFs and FFs for
arbitrary hadronic species (all unpolarised, all polarised, or partly
unpolarised and partly polarised) are convolved with partonic cross sections.
All of these distributions evolve differently: PDFs according to unpolarised
(polarised) space-like evolution equations; FFs according to
unpolarised (polarised) time-like evolution equations.

The \pineappl{\sc v1} data structure has been developed for an arbitrary number
of non-perturbative input distributions and convolutions, thus ensuring complete
flexibility and scalability of the framework. This may be required, {\it e.g.},
when modelling multi-parton scattering processes. In most cases, however,
the computation of physical observables require only a few convolutions.
For the sake of the description of the \pineappl{\sc v1} data structure, let us
consider the inclusive production of a neutral pion in an unpolarised
proton--proton collision, $p(P_A)\, p(P_B)\to \pi^0(P_h) + X$,
with $P_A$, $P_B$, and $P_h$ the four-momenta of the two incoming protons and
of the outgoing pion, respectively. At leading twist, the corresponding
cross section factorises as
\begin{align}
  \frac{d\sigma}{d\mathcal{O}}(\mathcal{O}, Q^2, \mu^2_R, \mu^2_F, \mu^2_f)
  & =
  \sum_{a,b,c}\int dx_1\, dx_2\, dz\,
  f_a\!\left( x_1,\mu_F^2\right)
  f_b\!\left( x_2,\mu_F^2\right)
  D_c^{\pi}\!\left( z, \mu_f^2\right)
  \nonumber\\
  & \times
  \frac{d\hat\sigma_{ab\to c}}{d\mathcal{O}}\left(\mathcal{O},Q^2,x_1,x_2,z,\xi_F^2,
  \xi_R^2, \xi_f^2 \right)\,
  \label{eq:cross-section-factorisation}
\end{align}
where, similarly to Eq.~\eqref{eq:pp_incl}, $\mathcal{O}$ is a set of kinematic
variables in which the cross section is differential, $Q^2$ is the
characteristic scale of the process ({\it e.g.}, the transverse momentum of the
outgoing pion), $\mu_R$ is the renormalisation scale,  $\mu_F$ and $\mu_f$
($\xi_f^2=\mu_f^2/Q^2$) are the PDF and FF factorisation scales, respectively,
$f_a$ and $f_b$ are the proton PDFs, $D_c^\pi$ is the pion FF, $x_1$ and $x_2$
are the fractions of the initial-state proton momenta carried by the partons
involved in the hard scattering, $z$ is the momentum fraction of the
fragmenting parton carried by the produced hadron, $\hat\sigma_{ab\to c}$ is the
partonic cross section, and the indexes $a$, $b$, and $c$ denote all the active
partons at a given scale. In principle, the cross section can be further
integrated in bins (of some) of the kinematic variables $\mathcal{O}$ or $Q^2$,
in which case an additional integration, over the bin limits, appears in
Eq.~\eqref{eq:cross-section-factorisation}. The cross section will then
retain an explicit dependence on the integration limits, which we omit in
Eq.~\eqref{eq:cross-section-factorisation}.

\pineappl\ deals with the partonic cross section $\hat\sigma$ in
Eq.~\eqref{eq:cross-section-factorisation}, which can be rewritten
as an expansion in powers of the strong coupling $\alpha_s$, the
electroweak coupling $\alpha$, and the scale logarithms as in
Eq.~\eqref{eq:pp_incl_pxs}
\begin{align}
  \frac{d\hat{\sigma}_{ab \to c}}{d\mathcal{O}}
  \left(\mathcal{O}, Q^2, x_1, x_2, z,
  \xi_R^2,\xi_F^2,\xi_f^2 \right)
  & = \sum_{\mathclap{k, \ell, m, n, p}} \alpha_s^k
  \left(\mu_R^2\right) \alpha^\ell
  \log^m \left( \xi_R^2 \right)
  \log^n \left(\xi_F^2 \right)
  \log^p \left(\xi_f^2 \right)
  \nonumber \\
  & \times W^{(k, \ell, m, n, p)}_{ab \to c}
  \left( \mathcal{O}, Q^2, x_1, x_2, z\right)\,.
  \label{eq:power_expansion}
\end{align}
In calculations in which the phase-space integration is performed using Monte
Carlo techniques, finite statistics does not allow for the exact reconstruction
of the dependence of the cross section on the kinematic variables $\mathcal{O}$.
In this case, it is sufficient to approximate the derivative with a piece-wise
constant function, such that
\begin{equation}
  W_{ab\to c}^{(k,l,m,n,p)}
  \left( \mathcal{O}, Q^2, x_1, x_2, z\right)
  \approx
  \sum_{o=1}^M\frac{\Theta(\mathcal{O}_o^{\rm min}\leq \mathcal{O}<\mathcal{O}_o^{\rm max})}{\mathcal{O}_o^{\rm max}-\mathcal{O}_o^{\rm min}}
  w_{ab\to c}^{(k,l,m,n,p,o)}\left(Q^2, x_1, x_2, z\right)\,.
  \label{eq:piecewise}
\end{equation}

The quantity $w$ is what \pineappl\ encodes in a grid representation.
As mentioned in Sect.~\ref{subsec:general_principles}, a
Monte Carlo set of $N$ tuples
$\left\{ Q^2_i, x_1^i, x_2^i, z^i, w_{ab\to c}^{(k,l,m,n,p,o)}\left(Q^2_i, x_1^i, x_2^i, z^i\right) \right\}_{i=1}^N$ could be associated to neighbouring points of a
suitably chosen grid that interpolates over them. In this case, the
convolution integral in Eq.~\eqref{eq:cross-section-factorisation} reduces
to the sum over grid points of the products of PDFs, FFs, and interpolation
coefficients of the weights. As also noted in
Sect.~\ref{subsec:general_principles}, this strategy is inefficient because
it does not generalise efficiently to an arbitrary dimension of the tuple (in
case of more convolutions), and is prone to wasting memory space by retaining
portions of the phase space that are devoid of Monte Carlo events.

To enable an arbitrary number of convolutions, \pineappl{\sc v1} generalises
the array that stores the interpolation coefficients, such as
$A_{j_{Q^2}, j_{x_1}, j_{x_2}, j_z}^{(k,l,m,n,p,o,a,b,c)}$ for the weights in
Eq.~\eqref{eq:piecewise}, to an arbitrary number of dimensions. In more
general processes, indeed, the number of indices of the form $j_{x_1}$, $j_{x_2}$,
$j_z$, $\ldots$ is not fixed, but is determined by the number of convolution
variables. The internal container used to store these coefficients in
\pineappl{\sc v1} is called \texttt{PackedArray}. Conceptually, it represents
an $n$-dimensional array, with one axis for each interpolation variable.
Internally, however, this $n$-dimensional array is linearised into a
one-dimensional array. A multi-index
\begin{equation}
  (j_0,j_1,\ldots,j_{n-1})
\end{equation}
is mapped to a single integer index, for instance in row-major order,
\begin{equation}
  J = j_{n-1}
      + N_{n-1} j_{n-2}
      + N_{n-1}N_{n-2} j_{n-3}
      + \cdots ,
\end{equation}
where $N_r$ is the number of interpolation nodes along dimension $r$.
Every point of the conceptual multidimensional grid is assigned a
unique position in a one-dimensional storage array.

A dense storage of this one-dimensional array would still be inefficient,
because most interpolation nodes are never filled. This happens because
phase-space cuts and kinematic constraints populate only a restricted region of
the full interpolation hypercube. Since this region is usually not known
exactly before running the Monte Carlo generator, it is in general not possible
to restrict the grid to the relevant phase-space region beforehand.
The {\tt PackedArray} is therefore sparse:
entries that are identically zero are not stored. Moreover, to avoid storing the
position of every non-zero entry separately, consecutive non-zero entries in the
linearised array are grouped together. Each group stores its starting position,
its length, and the corresponding sequence of floating-point coefficients. In
this way, the memory cost scales with the actually populated part of the
interpolation grid, up to a small bookkeeping overhead, rather than with the
size of the full dense hypercube.

\begin{figure*}[!t]
  \centering
  \includegraphics[width=\textwidth]{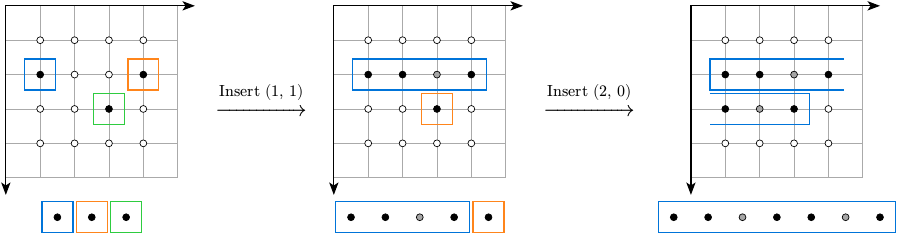}\\
  \caption{Visualisation of filling the \texttt{PackedArray}, here in the 2D
    4$\times$4 case. We show how the conceptual array (upper row) and the actual
    stored data (lower row) change when filling two elements at $(1, 1)$ and
    $(2, 0)$. Non-zero elements are indicated in black, explicitly stored
    zeros in grey, and implicit (non-stored) zeros in white. Elements that are
    grouped together are surrounded by a coloured rectangle.}
  \label{fig:packed-array}
\end{figure*}

\begin{figure}[!t]
	\centering
	\includegraphics[width=\textwidth]{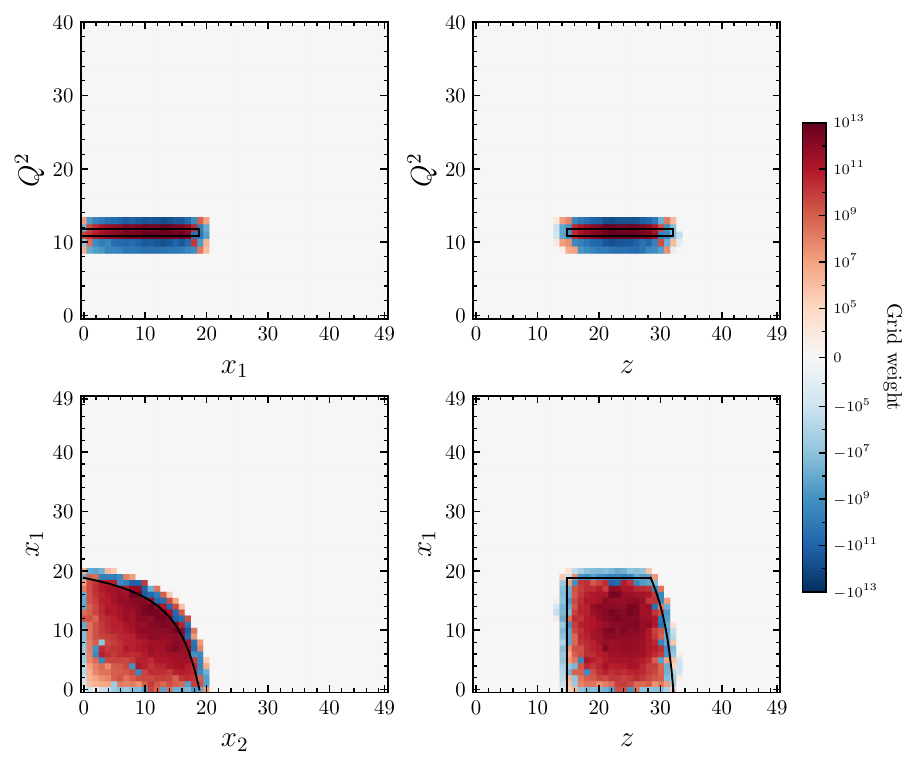}\\
    \caption{Subgrid contents of one of the bins in a single-inclusive hadron
      production grid, projected onto pairs of interpolation axes. The coloured
      pixels correspond to populated interpolation nodes, with the colour scale
      indicating the magnitude of the stored interpolation coefficients on a
      logarithmic scale. The solid black curves denote the analytic integration
      boundaries.}
	\label{fig:grid-sparsity}
\end{figure}

Figure~\ref{fig:packed-array} illustrates this mechanism with a two-dimensional
$4\times4$ toy array. The two-dimensional case is used only for visualisation;
the same procedure applies to any number of dimensions. The upper row shows the
conceptual array, while the lower row shows the linearised storage.
Black points denote non-zero interpolation coefficients, white points denote
zero coefficients that are not stored, and grey points denote zeros that are
stored explicitly. Initially the three non-zero coefficients are isolated and
therefore form three separate groups. When the coefficient at multi-index
$(1,1)$ is filled, it connects previously separated entries in the linearised
representation. It can then be cheaper to store the intervening zero
explicitly, shown in grey, and merge the neighbouring entries into a single
group, because storing one extra floating-point value may cost less memory
than storing the start position and length of an additional group.
The final insertion at $(2,0)$ also makes clear that adjacency is defined after
linearisation: entries that belong to different rows of the conceptual
two-dimensional array can be consecutive in the internal one-dimensional
storage.

To illustrate how storing only the positions and sizes of groups, rather than
the positions of all individual elements, leads to a significant reduction of
the memory footprint, Fig.~\ref{fig:grid-sparsity} shows the content of a
representative subgrid for one bin of a single-inclusive hadron production grid,
projected onto pairs of interpolation axes. The populated interpolation nodes
are displayed as coloured pixels, with the colour scale indicating the
magnitude of the stored interpolation coefficients on a logarithmic scale.
The solid black lines denote the analytic integration boundaries that constrain
the physically accessible phase space. The projections illustrate how only a
restricted region of the interpolation hypercube is populated by Monte Carlo
events, reflecting the underlying kinematic correlations among the convolution
variables $x_1$, $x_2$, and $z$. This sparsity motivates the
\texttt{PackedArray} representation introduced above, in which only the
populated regions of the multidimensional interpolation grid are retained in
memory.

\subsection{Scale functional forms and scale variation prescriptions}
\label{subsec:scale_variations}

Many processes of phenomenological interest are characterised by more than one
hard scale. In heavy-quark hadroproduction, for instance, both the transverse
momentum $p_T$ of the produced quark and its mass $M$ contribute to the
hardness of the interaction, and neither alone provides an unambiguous choice
for the physical scales $\mu_R$, $\mu_F$, and $\mu_f$. In such cases, the
physical scales must be constructed as some combination of the individual
kinematic scales, with different calculations in the literature adopting
different prescriptions.

\pineappl{\sc v1} accommodates this by allowing the grid to store two
independent kinematic scale variables $Q_1$ and $Q_2$ for each event, and
letting each of $\mu_R$, $\mu_F$, and $\mu_f$ be defined independently via
one of the functional forms collected in Table~\ref{tab:scale-forms}. The
three scales may thus assume entirely different forms, reflecting, for instance,
a calculation in which $\mu_R$ and $\mu_F$ are set by different combinations
of $Q_1$ and $Q_2$.

\begin{table}[!t]
  \centering
  \renewcommand{\arraystretch}{1.4}
  \begin{tabularx}{\textwidth}{XccXc}
    \toprule
      functional form & scale choice
      & & functional form & scale choice\\
      \midrule
      quadratic sum
      & $Q_1^2+Q_2^2$
      & & minimum
      & $\min(Q_1, Q_2)^2$\\
      quadratic mean
      & $\frac{1}{2}(Q_1^2+Q_2^2)$
      & & product
      & $Q_1^2Q_2^2$\\
      quadratic sum/4
      & $\frac{1}{4}(Q_1^2+Q_2^2)$
      & & fourth-power mean
      & $\sqrt{Q_1^4+Q_2^4}$\\
      linear sum
      & $(Q_1+Q_2)^2$
      & & weighted average
      & $\dfrac{Q_1^4+Q_2^4}{Q_1^2+Q_2^2}$\\
      linear mean
      & $\frac{1}{4}(Q_1+Q_2)^2$
      & & shifted quadratic (1/2)
      & $Q_2^2+\frac{1}{2}Q_1^2$\\
      maximum
      & $\max(Q_1,Q_2)^2$
      & & shifted quadratic (1/4)
      & $Q_2^2+\frac{1}{4}Q_1^2$\\
    \bottomrule
  \end{tabularx}
  \caption{Functional forms available in \pineappl{\sc v1} for constructing
    each of the physical scales $\mu_R$, $\mu_F$, and $\mu_f$ from one or
    two kinematic scale variables $Q_1$ and $Q_2$ stored in the interpolation
    grid. The single scale $Q^2_i$ is not shown. All three scales can
    independently adopt different forms.}
  \label{tab:scale-forms}
\end{table}

On the other hand, one of the main advantages of interpolation grids is the
ability to quickly evaluate physical observables upon changes of the input
distribution functions or of the logarithmic scale contributions. The latter
are in particular useful to assess missing higher-order uncertainties by means
of scale variations. \pineappl\ allows one to store the dependence on the
renormalisation $\mu_R$ and factorisation scales $\mu_F$ and $\mu_f$ in the
PDFs and FFs directly in the grids. This feature enables scale variations
through any linear transformation of the original scale used for the grid
generation in a straightforward way.

Scale variations are conventionally performed by varying
$\xi_t=\sqrt{\mu_t^2/Q^2}$ (for $t=R,F,f$) by factors of $(1/2, 2)$ with
respect to the central scale $Q$. There exist various prescriptions for scale
variations. When three scales are involved, the most commonly used ones are the
following.

\begin{figure*}[tb]
  \centering
  \includegraphics[width=0.32\textwidth]{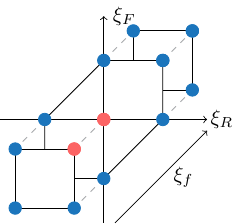}
  \includegraphics[width=0.32\textwidth]{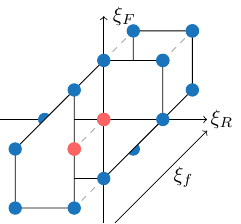}
  \includegraphics[width=0.32\textwidth]{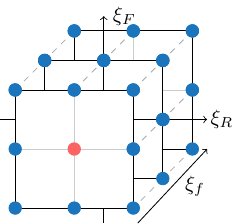}
  \caption{The areas spanned by the variation of the renormalisation scale $\xi_R$,
    and factorisation scales $\xi_F$ and $\xi_f$ for the 15-point
    (left), 17-point (middle), and 27-point (right) prescriptions.
    The central scales $(\xi_R, \xi_F) = (1, 1)$ are marked by the red points
    while the different layers represent the fragmentation scale
    $\xi_f = 1/2, 1, 2$.}
  \label{fig:scale-variations}
\end{figure*}

\begin{itemize}

\item \textbf{7-point prescription:} the renormalisation scale $\mu_R$ and
  factorisation scale $\mu_F$ are varied independently while the factorisation
  scales $\mu_F$ and $\mu_f$ are correlated and set to be
  $\mu_F=\mu_f$~\cite{Caletti:2024xaw, Bonino:2024wgg}. This variation is
  exactly the same as the 7-point prescription in the case of only two scales.

\item \textbf{15-point prescription:} the three scales $(\mu_R, \mu_F, \mu_f)$
  are varied independently probing all the combinations except for the extreme
  diagonal. These are combinations where two scales differ by a factor greater
  than 2, {\it i.e.}, when $\mu_i / \mu_j = 1/4$ or $4$ for any
  $i,j \in \lbrace R, F, f \rbrace$. This variation is the equivalent of the
  two-scales 7-point variation in the three-scale case. The area spanned by
  such a variation is shown in the left panel of
  Fig.~\ref{fig:scale-variations}.

\item \textbf{17-point prescription:} the three scales are varied
  independently, omitting combinations in which $(\mu_R, \mu_F)$ or
  $(\mu_R, \mu_f)$ are pairwise scaled by a factor of two in the
  opposite directions~\cite{dEnterria:2013sgr,Banfi:2010xy}.
  The areas spanned by such a variation is shown in the middle panel of
  Fig.~\ref{fig:scale-variations}.

\item \textbf{27-point prescription:} the three scales are varied
  independently as in the 15- and 17-point prescriptions but includes all the
  $3^3 = 27$ combinations, covering the full cube as diagrammatically
  illustrated in the right panel of Fig.~\ref{fig:scale-variations}.

\end{itemize}
We will use the 15-point prescription to illustrate some phenomenological
applications in Sect.~\ref{sec:pheno}.

\section{Phenomenological applications of \pineappl\ grids}
\label{sec:pheno}

In this section, we demonstrate the new features of \pineappl\ by computing
predictions of cross sections corresponding to the inclusive production of an
identified hadron in proton--proton collisions and in SIDIS. These processes
have been measured by various experiments, that typically also allow for the
longitudinal polarisation of the initial-state protons or leptons.
The computation of the corresponding cross sections therefore combines up to
three different distributions, which all evolve differently: unpolarised and
polarised PDFs, and unpolarised FFs. For each process, we define the
physical observables of interest, present the set of measurements that we
consider, discuss the computational settings that we use, and comment
on some phenomenological implications.

\subsection{Single-inclusive hadron production in proton--proton collisions}
\label{subsec:pheno-pp}

The computation of cross sections for the single-inclusive production of a
hadron in proton--proton collisions requires the convolution of two PDFs, for
each of the nucleons in the initial state, and a FF, for the identified hadron
in the final state. Because measurements exist in which the two protons in the
initial state are either unpolarised or longitudinally polarised one needs to
consistently use unpolarised or polarised PDFs, thus combining (un)polarised
space-like evolution with unpolarised time-like evolution. The physical
observables that are typically measured are the Lorentz-invariant cross
section, in the unpolarised case, and the double-spin asymmetry, in the
polarised case. These are defined, respectively, as (modulo overall kinematic
factors)
\begin{equation}
  E\frac{d^3\sigma}{dp^3}(p,Q^2,\mu_R^2,\mu_F^2,\mu_f^2)
  \propto
  \sum_{a,b,c}
  f_a(x_1,\xi_F^2)
  \otimes
  f_b(x_2,\xi_F^2)
  \otimes
  D_c^h(z,\xi_f^2)
  \otimes
  \hat\sigma_{ab}^c(x_1, x_2, z, \xi^2_R,\xi^2_F,\xi^2_f)\,,
  \label{eq:Lorentz}
\end{equation}
\begin{equation}
  A_{LL} = \frac{Ed^3\Delta\sigma/dp^3}{Ed^3\sigma/dp^3}\,,
  \label{eq:double-spin}
\end{equation}
where $Ed^3\Delta\sigma/dp^3$ is obtained from Eq.~\eqref{eq:Lorentz} by
replacing the unpolarised PDFs $f_{a,b}$ and partonic cross section
$\hat\sigma_{ab}^c$ with their polarised counterparts $\Delta f_{a,b}$ and
$\Delta\hat\sigma_{ab}^c$. The notation in Eq.~\eqref{eq:Lorentz} is the same as
in Eq.~\eqref{eq:cross-section-factorisation}, except for the fact that the
variable $\mathcal{O}$ is the momentum $p$ of the hadron in the final state
(with energy $E$) and the symbol $\otimes$ is used to denote the convolutional
product
\begin{equation}
  g(x)\otimes h(x)=\int_x^1\frac{dz}{z}g\left(\frac{x}{z} \right) h(z)\,.
  \label{eq:convolution}
\end{equation}

The leading-order contribution to the Lorentz-invariant cross section,
Eq.~\eqref{eq:Lorentz}, comes from partonic channels in which a gluon fragments
into a final-state hadron. Measurements of single-inclusive hadron production
in unpolarised proton--proton collisions are therefore key to probe the gluon
FF. On the other hand, the leading-order contribution to the double-spin
asymmetry, Eq.~\eqref{eq:double-spin}, comes from partonic channels with
initial-state polarised gluons. Measurements of the double-spin asymmetry are
therefore key to probe the polarised gluon PDF.

We consider the following measurements. For unpolarised proton--proton
collisions, we focus on neutral pion production Lorentz-invariant cross
sections measured at mid-rapidity by the STAR experiment at the BNL RHIC, and
by the ALICE experiment at the CERN LHC. Specifically, for STAR we select the
measurement performed at a centre-of-mass energy of
200~GeV~\cite{STAR:2009vxb}, whereas for ALICE we select the measurements
performed at centre-of-mass energies of
0.9~TeV~\cite{ALICE:2012wos}, 2.76~TeV~\cite{ALICE:2017nce},
7 TeV~\cite{ALICE:2012wos}, 8~TeV~\cite{ALICE:2017ryd}, and
13~TeV~\cite{ALICE:2020jsh}. They correspond, respectively, to luminosities of
3~pb$^{-1}$ (2005 run), 0.14~nb$^{-1}$, 0.52~nb$^{-1}$, 5.6~nb$^{-1}$,
1.25~nb$^{-1}$, and 1.1~nb$^{-1}$. All measurements are differential in the
transverse momentum of the neutral pion, $p_T$, and are integrated over the
rapidity range $|y|<1$ for STAR and $|y|<0.9$ for ALICE, except for the 13~TeV
measurement, where $|y|<0.5$. For polarised proton--proton collisions, we focus
on neutral pion production double-spin asymmetries measured at mid-rapidity
by the PHENIX experiment at the BNL RHIC. Specifically, we select measurements
at centre-of-mass energies of 200~\cite{PHENIX:2007kqm} and
510~GeV~\cite{PHENIX:2015fxo}. They correspond, respectively, to
luminosities of 2.5~pb$^{-1}$ (2005 run), and 108~pb$^{-1}$ (2013 run). All
measurements are differential in the transverse momentum of the neutral pion,
$p_T$, and are integrated over the rapidity range $|y|<0.35$.

We compute theoretical predictions with a modified version of the code developed
in~\cite{deFlorian:2002az}, which we have interfaced to {\sc PineAPPL},
see~\cite{Cruz-Martinez:2025ahf}. The default accuracy of our computations is
NLO in the perturbative expansion of the strong coupling. We choose the central
scale $Q^2=\mu^2_R=\mu^2_F=\mu^2_f=p_T^2$. We estimate missing higher-order
uncertainties by means of scale variations, using the 15-point prescription,
as explained in Sect.~\ref{subsec:scale_variations}. In the case of
unpolarised proton--proton collisions, NNLO corrections have also been computed
very recently~\cite{Czakon:2025yti}, and have been delivered in the form of
$K$-factors~\cite{Czakon:K-factors}, although without scale variations.
These allow us to readily estimate the impact of higher-order
corrections on the central value of our predictions, but make the estimate of
the reduction of missing higher-order corrections through scale variations 
complicated. We use the {\sc NNPDF4.0} unpolarised proton PDF
set~\cite{NNPDF:2021njg}, the {\sc NNPDFpol2.0} polarised proton PDF
set~\cite{Cruz-Martinez:2025ahf}, and the {\sc NNFF1.0} neutral pion FF
set~\cite{Bertone:2017tyb}. We specifically use the baseline NLO and NNLO
sets, corresponding to a value of the strong coupling at the $Z$ pole mass
$\alpha_s(M_Z)=0.118$. We have checked that the loss in accuracy due to
the polynomial interpolation in the {\sc PineAPPL} grids is negligible in
comparison to the Monte Carlo integration uncertainty of the native results
obtained with the code of~\cite{deFlorian:2002az}. A systematic benchmark is
presented in Appendix~\ref{app:MC}.

\begin{figure}[!t]
  \centering
  \includegraphics[width=\textwidth]{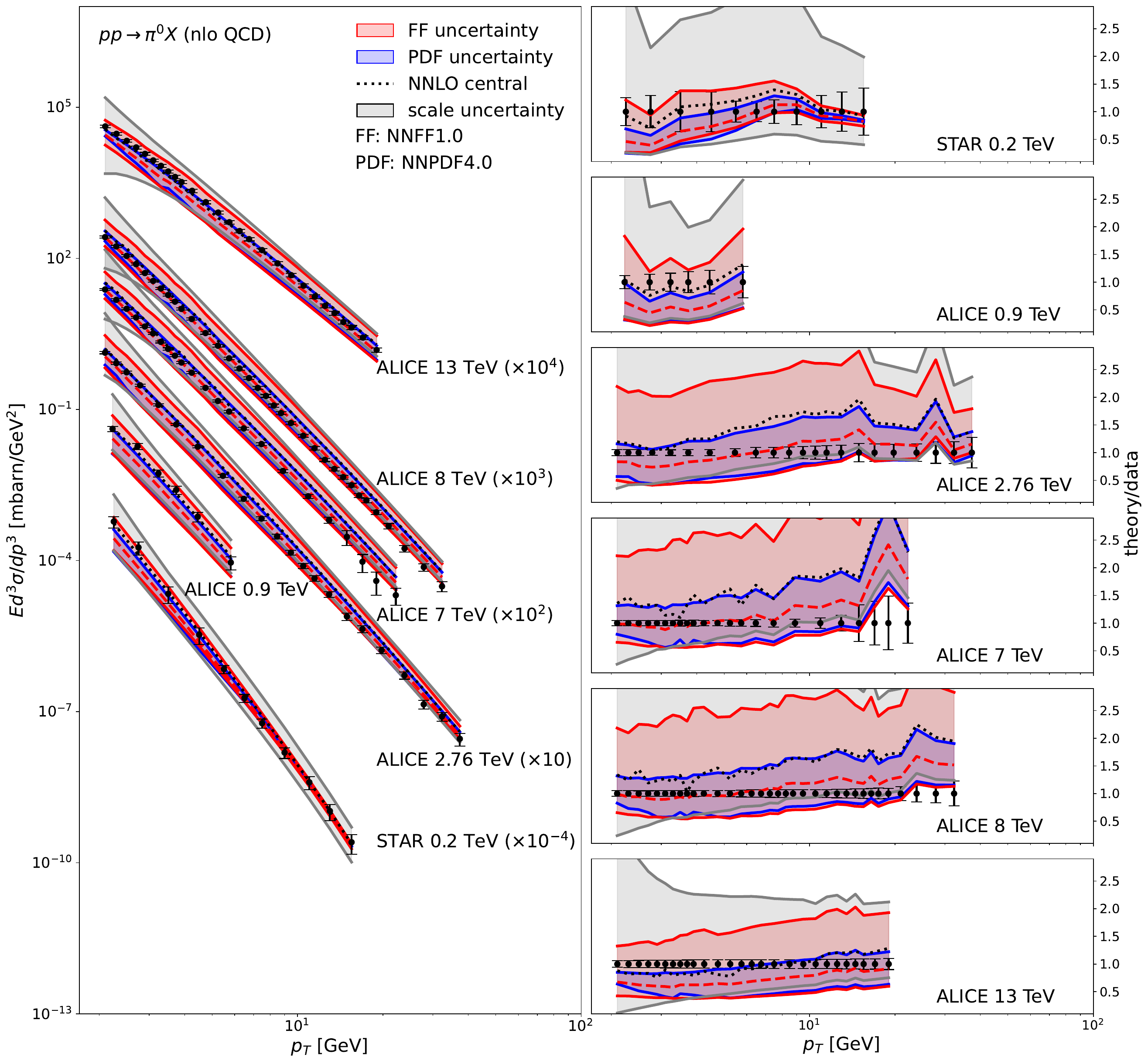}\\
  \caption{The Lorentz-invariant cross section, Eq.~\eqref{eq:Lorentz},
    for the inclusive production of a neutral pion in unpolarised proton--proton
    collisions, as a function of the transverse momentum of the pion.
    Measurements from STAR~\cite{STAR:2009vxb} and ALICE~\cite{ALICE:2012wos,
      ALICE:2017nce,ALICE:2012wos,ALICE:2017ryd,ALICE:2020jsh} experiments, at
    various centre-of-mass-energies, are compared to predictions, accurate to
    NLO in the strong coupling, obtained from \pineappl\ grids generated
    in turn with the code presented in~\cite{deFlorian:2002az,
      Cruz-Martinez:2025ahf}. The NNPDF4.0 PDFs and NNFF1.0 FFs are used as
    input. Uncertainty bands represent the 68\% confidence level PDF and FF
    uncertainties, and the scale uncertainty computed with the 15-point
    prescription. The central value of the NNLO prediction, computed with NNLO
    PDFs and FFs and supplemented with $K$-factors, is also shown.}
  \label{fig:pp_unp}
\end{figure}

\begin{figure}[!t]
  \centering
  \includegraphics[width=\textwidth]{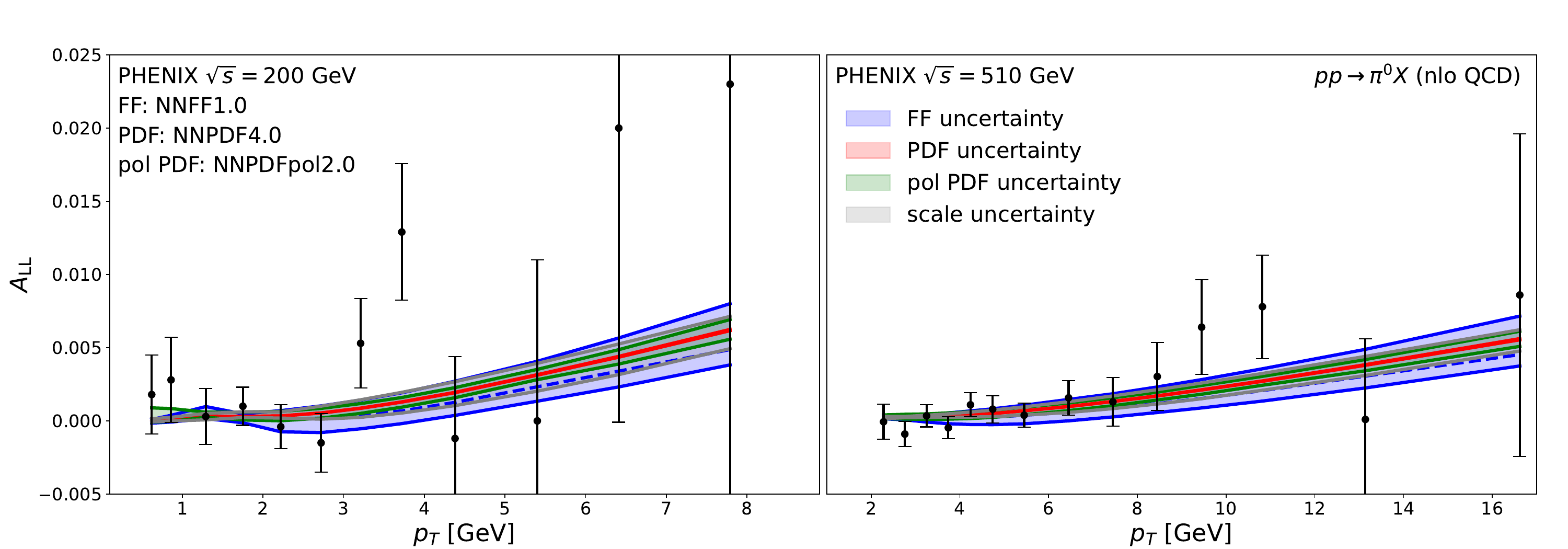}\\
  \caption{The double-spin asymmetry, Eq.~\eqref{eq:double-spin}, for the
    inclusive production of a neutral pion in polarised proton--proton
    collisions, as a function of the transverse momentum of the pion.
    Measurements from the PHENIX experiment at centre-of-mass energies of
    200~GeV~\cite{PHENIX:2007kqm} and 510~GeV~\cite{PHENIX:2015fxo}
    are compared to predictions, accurate to NLO in the strong coupling,
    obtained from \pineappl\ grids generated in turn with the code presented
    in~\cite{deFlorian:2002az,Cruz-Martinez:2025ahf}. The NNPDF4.0 PDFs,
    NNPDFpol2.0 polarised PDFs, and NNFF1.0 FFs are used as input. Uncertainty
    bands represent the 68\% confidence level PDF and FF uncertainties, and the
    scale uncertainty computed with the 15-point prescription.}
  \label{fig:ALL_PHENIX}
\end{figure}

We now turn to the comparisons between the experimental measurements described
above and the corresponding theoretical predictions. In Fig.~\ref{fig:pp_unp},
we display such a comparison for the unpolarised case, and in
Fig.~\ref{fig:ALL_PHENIX} for the polarised case. We show separately the PDF,
FF, and scale uncertainties. In the former case, we also show the central value
of the NNLO predictions, computed with NNLO PDF and FF sets and supplemented
with $K$-factors. In the latter case, we also display the polarised PDF
uncertainty. In all cases, PDF and FF uncertainty bands correspond to the 68\%
confidence level, computed over the Monte Carlo replicas in the corresponding
parton sets.

In the unpolarised case, we observe that the NLO scale uncertainty is the
largest, and, as expected, it increases as $p_T$ and the centre-of-mass
energy decrease. The second largest uncertainty is the FF uncertainty, which is
two to three times larger than the PDF uncertainty. The agreement between the
measurements and the theoretical predictions is good, within the PDF, FF, and
scale uncertainties. This fact is worth noting, because the input NNPDF4.0 and
NNFF1.0 PDF and FF sets do not include any of these measurements in the
corresponding determinations. In this case, they therefore generalise very well
to unseen data. The NNLO corrections are large, though consistent with the NLO
scale variations: $K$-factors are of the order of 5-15\%, with an increase as
both $p_T$ and the centre-of-mass energy decrease, whereas the impact of
switching from NLO to NNLO distributions accounts for another 3-6\%.
However, they do not always bring the central value of the
predictions closer to the experimental data. Overall, the uncertainty on
theoretical predictions is much larger than the uncertainty on the measurements.
This fact suggests that, once these will be included in a determination of
PDFs and FFs, they will potentially have a significant impact. The large NLO
scale uncertainties and NNLO corrections, however, suggest that this will be
useful only if the full NNLO computation is retained.

In the polarised case, the situation is comparatively simpler. The largest
uncertainty is the FF uncertainty, then come the polarised PDF and scale
uncertainties, which are half of that, and finally the unpolarised PDF
uncertainty, which is negligible. The agreement between the measurements and
the theoretical predictions is good, within the rather large uncertainty of
the former. Also in this case, this fact is remarkable, given that the 
input NNPDF4.0, NNPDFpol2.0, and NNFF1.0 sets do not include any of
these measurements in the corresponding determinations. This is yet another
evidence of the generalisation power of the used parton sets. The fact that NLO
scale variations are not dominant suggests that the need for NNLO corrections
is perhaps less compelling than in the unpolarised case. We finally note that
the experimental uncertainties are larger than or of the same order as the
uncertainties on the predictions. We therefore expect the impact of these
measurements in a determination of PDFs and/or FFs to be moderate.

\subsection{Single-inclusive hadron production in SIDIS}
\label{subsec:pheno-sidis}

The computation of cross sections for the single-inclusive production of a
hadron in lepton-nucleon scattering requires the convolution of a PDF and
of a FF. There exist measurements for SIDIS multiplicities,
{\it i.e.}, the ratio of the unpolarised SIDIS to DIS cross sections,
and the SIDIS double-spin asymmetry, {\it i.e.}, the ratio of the polarised to
unpolarised SIDIS cross sections. Neglecting kinematic corrections, this is
well approximated by the ratio of the polarised to unpolarised photoabsorption
asymmetries, that read respectively as (modulo overall kinematic factors)
\begin{align}
  g_1^h(x,z,Q^2,\mu_R^2,\mu_F^2,\mu_f^2)
  & \propto
  \sum_{a,c}
  \Delta f_a(x,\xi_F^2)
  \otimes
  D_c^h(z,\xi_f^2)
  \otimes
  \Delta\hat\sigma_a^c(x,z,\xi_R^2,\xi_F^2,\xi_f^2)\,,
  \label{eq:g1h}\\
  F_1^h(x,z,Q^2,\mu_R^2,\mu_F^2,\mu_f^2)
  & \propto
  \sum_{a,c}
  f_a(x,\xi_F^2)
  \otimes
  D_c^h(z,\xi_f^2)
  \otimes
  \hat\sigma_a^c(x,z,\xi_R^2,\xi_F^2,\xi_f^2)\,,
  \label{eq:F1h}  
\end{align}
hence the double-spin asymmetry reads as
\begin{equation}
  A_{LL}\approx A_1^h=\frac{g_1^h}{F_1^h}\,.
  \label{eq:double-spin-SIDIS}
\end{equation}
The notation used in Eqs.~\eqref{eq:g1h}--\eqref{eq:F1h} is the same as in
Eqs.~\eqref{eq:cross-section-factorisation}--\eqref{eq:Lorentz}. The observable
$\mathcal{O}$ is represented by the variables $x$ and $z$, {\it i.e.}, the
momentum fraction carried by the parton out of the initial-state hadron, and
the momentum fraction carried out by the observed final-state hadron respectively. The
leading-order contribution to the double-spin asymmetry,
Eq.~\eqref{eq:double-spin-SIDIS}, comes from partonic channels with quarks
and antiquarks in the initial and final states. Measurements of $A_{LL}$ are
therefore sensitive to the flavour and anti-flavour decomposition of PDFs and
FFs.

We consider measurements of the SIDIS double-spin asymmetry performed by the
HERMES experiment~\cite{HERMES:2018awh} during the 1996--2000 running period,
in which the lepton beam was longitudinally polarised with an energy of
27.6~GeV. We restrict ourselves to the subset corresponding to the production
of a positively charged  pion in electron-deuteron scattering.
The asymmetry is reported in bins of $x$, $z$, and $Q^2$.
We compute theoretical predictions with a preliminary version of Virtual
Hadron Factory (vhf)~\cite{Sharma:vhf}, which is interfaced to \pineappl, at NLO and NNLO
accuracy in the strong coupling. We choose the central scale
$\mu_R^2=\mu_F^2=\mu_f^2=Q^2$, and we estimate missing higher-order uncertainties
by means of scale variations, using the 15-point prescription. The input
unpolarised and polarised PDFs and FFs are as in Sect.~\ref{subsec:pheno-pp},
namely we use NNPDF4.0, NNPDFpol2.0, and NNFF1.0, consistently at NLO or NNLO.
Results for additional HERMES bins are displayed in Sect.~5.2
of~\cite{Cruz-Martinez:2025ahf}, whereas a survey of other results,
including for COMPASS~\cite{COMPASS:2010hwr} and for projected
EIC~\cite{AbdulKhalek:2021gbh} measurements, will be presented
elsewhere~\cite{PineAPFEL:vhf}.

\begin{figure}[!t]
  \centering
  \includegraphics[width=\textwidth]{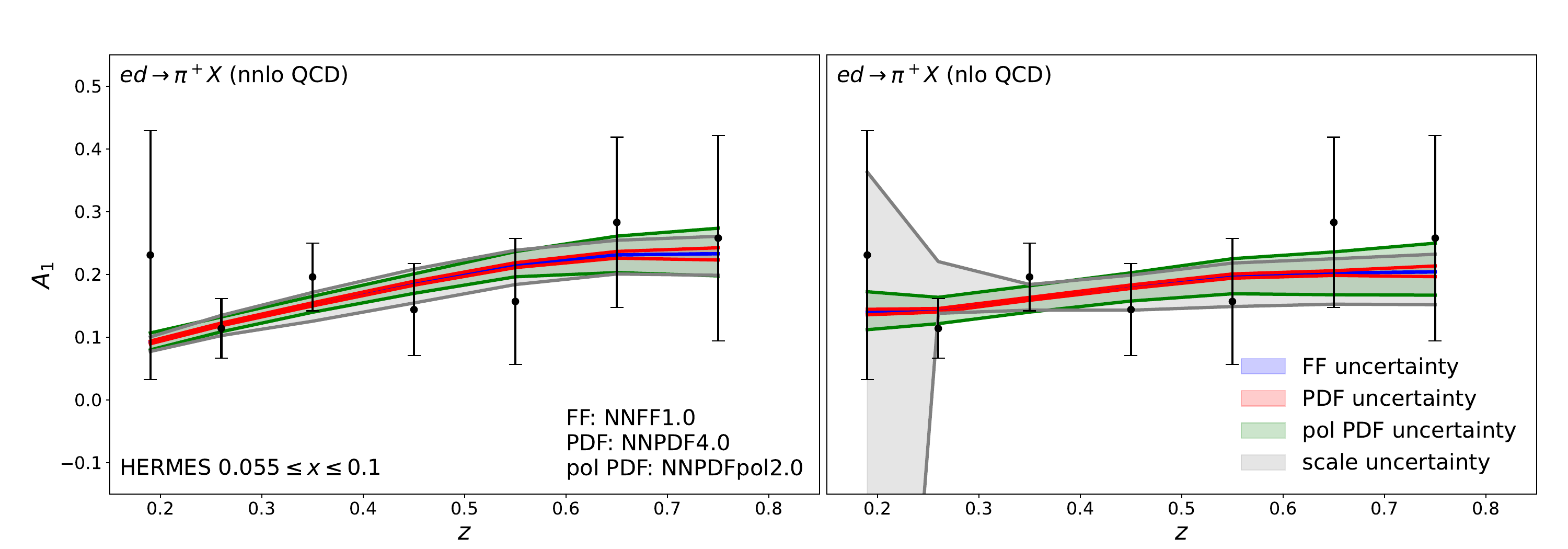}\\
  \caption{The double-spin asymmetry, Eq.~\eqref{eq:double-spin-SIDIS},
    for the inclusive production of a positively charged pion in polarised
    SIDIS, as a function of $z$ in a fixed bin of $x$. Measurements from the
    HERMES experiment~\cite{HERMES:2018awh} are compared to predictions,
    accurate to NLO and NNLO in the strong coupling, obtained from \pineappl\
    grids generated in turn with the code of~\cite{Sharma:vhf}.
    The NNPDF4.0 PDFs, NNPDFpol2.0 polarised PDFs, and NNFF1.0 FFs are used as
    input. Uncertainty bands represent the 68\% confidence level PDF and FF
    uncertainties, and the scale uncertainty computed with the 15-point
    prescription.}
  \label{fig:ALL_HERMES}
\end{figure}

In Fig.~\ref{fig:ALL_HERMES} we compare the HERMES experimental data, for the
bin $0.055\leq x \leq 0.1$, to our NLO and NNLO theoretical predictions.
We show separately the PDF (both unpolarised and polarised), FF, and scale
uncertainties. The PDF and FF uncertainties correspond to 68\% confidence
level bands. In this case, we observe that the dominant uncertainties are
the polarised PDF and scale uncertainties, which are similar in size. The
FF and PDF uncertainties are comparatively negligible. We observe, as expected,
a reduction of the scale uncertainty when moving from NLO to NNLO. The
agreement between the measurements and the theoretical predictions is good,
within the rather large uncertainty of the former. Also in this case, this
testifies the excellent generalisation power of the input NNPDF4.0,
NNPDFpol2.0, and NNFF1.0 sets to yet another set of unseen data. The HERMES
measurements are indeed not included in the determination of any of the used
parton sets. We finally note that the experimental uncertainties are larger
than or of the same order as the uncertainties on the predictions. We therefore
expect the impact of these measurements in a determination of PDFs and/or
FFs to be moderate.

\section{Conclusions and outlook}
\label{sec:conclusions}

We have presented \pineappl{\sc v1}, a major extension of the
\pineappl\ library for the fast evaluation of high-energy scattering
observables. \pineappl{\sc v1} preserves backward compatibility with
\pineappl{\sc v0}. The main new feature of this release is the support for
multiple convolutions with non-perturbative input distributions. This lifts
the restriction, common to previous fast-interpolation grid implementations, to
processes with two incoming hadronic distributions, and makes it possible
to describe observables that involve identified hadrons in the final state,
polarised initial states, or more general hadronic configurations.
New to \pineappl{\sc v1} is also the possibility to store two independent
kinematic scale variables in the grid, and to choose among a set of functional
forms that combine them. This is useful in processes that are characterised
by more than one scale, and for which the renormalisation and
factorisation scales are typically chosen as an analytic expression of two
hard scales. Renormalisation and factorisation scales, separately in the
PDFs and in the FFs, can be chosen and varied independently.

The extension required a corresponding generalisation of the grid
representation. We have reviewed the interpolation strategy already used in
the previous version of \pineappl, in which Monte Carlo phase-space weights are
not stored as event tuples, but are instead projected onto interpolation nodes.
In \pineappl{\sc v1}, the array of interpolation coefficients is promoted to an
arbitrary number of dimensions, one for each interpolation variable. The
resulting multi-dimensional array is stored internally through the
{\tt PackedArray} data structure, a sparse linearised representation that
stores only the populated regions of the interpolation hypercube and groups
neighbouring entries to reduce bookkeeping overhead. This provides a
memory-efficient representation whose dimensionality is
determined by the process under consideration rather than by the grid format.

We have also described how the new framework treats different kinds of
non-perturbative distributions. Each convolution can be associated with its own
hadron species and with its own type, distinguishing between unpolarised and
polarised distributions, and between space-like and time-like evolution. In this
way, a single grid can consistently combine, for instance, unpolarised PDFs,
polarised PDFs, and FFs. The dependence on the renormalisation scale and on the
initial- and final-state factorisation scales is encoded in the grids, allowing
for efficient a posteriori scale variations. In particular, we have discussed
standard prescriptions for simultaneous variations of the three scales that
arise when PDFs and FFs enter the same observable.

The new capabilities have been illustrated with two representative classes of
processes involving identified hadrons. First, we have considered
single-inclusive neutral pion production in proton--proton collisions, both in
the unpolarised and polarised cases. For unpolarised collisions, we compared
predictions obtained from \pineappl\ grids with STAR and ALICE measurements of
Lorentz-invariant cross sections over a wide range of centre-of-mass energies.
We found good agreement within the combined PDF, FF, and scale uncertainties.
The NLO scale uncertainty is the dominant contribution to the theoretical
uncertainty, while FF uncertainties are larger than PDF uncertainties. The
available NNLO corrections, applied through $K$-factors, are sizeable but
remain compatible with the NLO scale variation band. These results suggest that
such measurements have the potential to constrain FFs and PDFs, provided that
the perturbative description is retained at sufficiently high accuracy.
For polarised proton--proton collisions, we considered PHENIX measurements of
the double-spin asymmetry for neutral-pion production. In this case the dominant
uncertainty arises from the FFs, followed by the polarised PDFs and scale
variations, while the unpolarised PDF uncertainty is negligible. The predictions
are compatible with the data within the current experimental uncertainties. The
relative size of the uncertainty components indicates that these measurements
can provide useful information on the polarised gluon distribution and on the
FFs, although their impact in a global determination is expected to be moderate
with the present experimental precision. Second, we have studied SIDIS with an
identified positively charged pion in the final state. We compared NLO and NNLO
predictions for the HERMES double-spin asymmetry with the corresponding
experimental measurements. The predictions show good agreement with the data,
and the reduction of the scale uncertainty from NLO to NNLO confirms the
expected improvement in perturbative stability.
In the kinematic bin considered, the dominant theory uncertainties are the
polarised PDF and scale uncertainties, while the unpolarised PDF and FF
uncertainties are much smaller.

The developments presented here open the way to using fast-interpolation grids
for a much broader class of observables than previously possible. Immediate
applications include global determinations of unpolarised and polarised PDFs,
FFs, and possibly their simultaneous extraction from processes involving both
initial- and final-state hadrons. More generally, the arbitrary convolution
framework can be used for phenomenological studies of semi-inclusive processes
at RHIC, the LHC, the forthcoming EIC, and future collider
facilities. It also provides the technical basis for extending fast-grid methods
to more differential observables, multi-hadron final states, and processes in
which several non-perturbative distributions enter the same factorisation
formula. Future work will focus on interfacing additional higher-order
calculations to \pineappl{\sc v1}, extending the library of publicly available
grids, and further developing the command-line and programming interfaces for
use in precision QCD phenomenology.

\section*{Acknowledgments}

We thank Valerio Bertone, Felix Hekhorn, and Alex Huss for valuable discussions
regarding several features of PineAPPL, and for reporting bugs that significantly
improved the user experience.
E.R.N. was supported by the Italian Ministry of University and Research (MUR)
through the “Rita Levi-Montalcini” Program.
During the early stage of this work, T.R.R. was partially supported by an
Accelerating Scientific Discoveries (ASDI2021) with the grant from the Netherlands
eScience Center (NLeSC), grant number 027.020.G05. The work of T.R.R. is supported
by the French Agence Nationale de la Recherche (ANR) via the grant ANR-20-CE31-0015
(“PrecisOnium”) and was also partly supported by the French CNRS via the COPIN-IN2P3
bilateral agreement and via the IN2P3 project “QCDFactorisation@NLO”.
The work of T.J. and J.W. was supported by the BMBF under contract 05P24PMA.

\appendix
\section{Interpolation accuracy in \pineappl\ grids}
\label{app:MC}

In this Appendix, we check that the interpolation performed on the
\pineappl\ grid does not result in a loss of accuracy with respect to the
original Monte Carlo computation. We specifically focus on the physical cross
sections considered in Sect.~\ref{subsec:pheno-pp} for single-inclusive neutral
pion production in proton--proton collisions. We compare predictions computed
in two ways. First, by running the unmodified version of the code
of~\cite{Jager:2002xm} in which the convolution between PDFs/FFs and matrix
elements is performed at every step of the Monte Carlo integration. Second, by
convolving the PDFs/FFs with the grids generated with a version of the same
code, modified with an interface to \pineappl~\cite{Cruz-Martinez:2025ahf}.
We note that the same exercise cannot be performed with the SIDIS asymmetry
discussed in Sect.~\ref{subsec:pheno-sidis}, the reason being that the
code used in that case~\cite{Sharma:vhf} already performs its own interpolation
on subgrids, that are then collated when constructing a \pineappl\ grid.
Therefore, by construction, being there not an interpolation on the \pineappl\
side, there is not a loss in accuracy due to \pineappl.

\begin{figure}[!t]
  \centering
  \includegraphics[width=\textwidth]{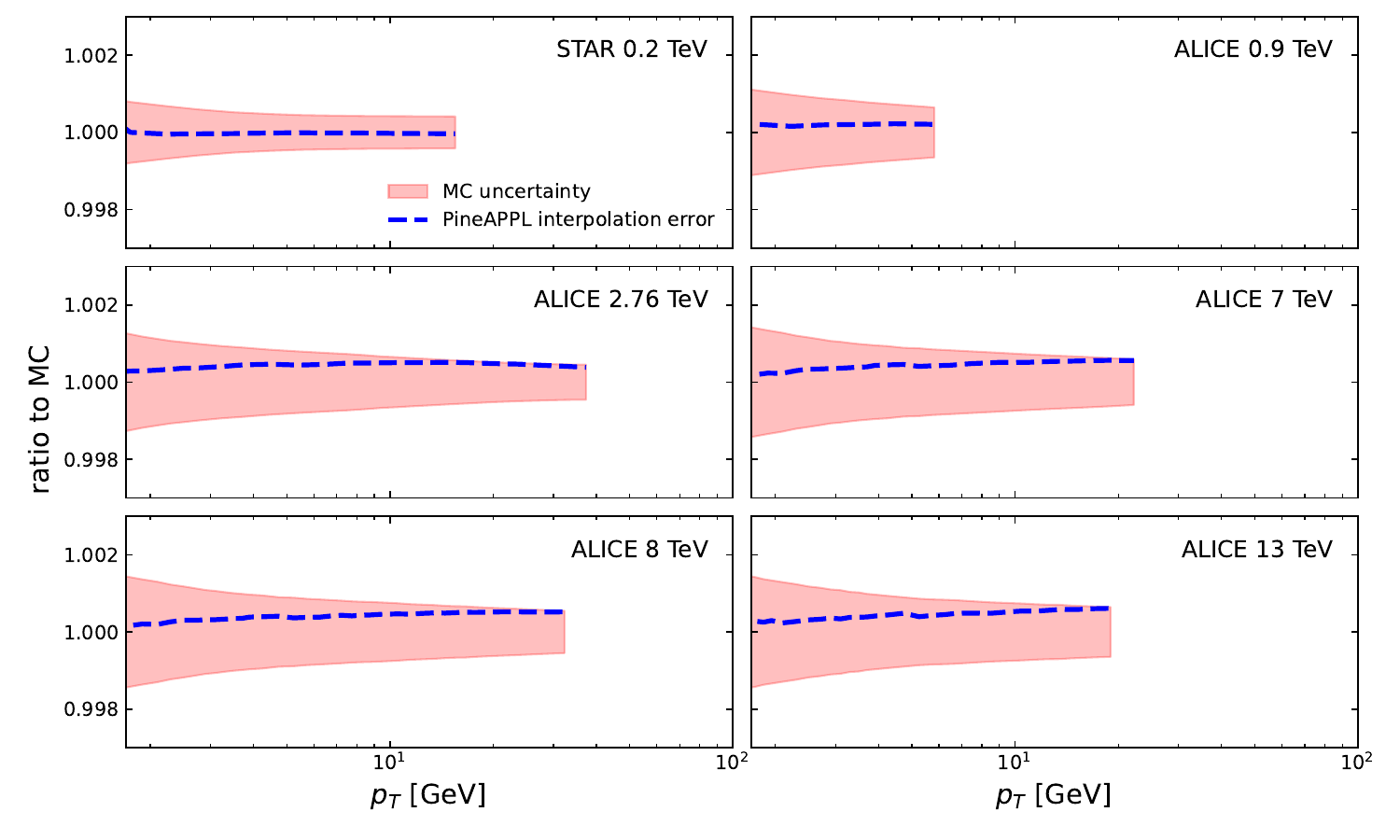}\\
  \caption{The relative Monte Carlo uncertainty on the predictions for the
    Lorentz-invariant cross section corresponding to the measurements
    reported in Sect.~\ref{subsec:pheno-pp}~\cite{STAR:2009vxb,ALICE:2012wos,
      ALICE:2017nce,ALICE:2012wos,ALICE:2017ryd,ALICE:2020jsh},
    compared to the \pineappl\ relative interpolation error.}
  \label{fig:unp_validation}
\end{figure}

\begin{figure}[!t]
  \centering
  \includegraphics[width=\textwidth]{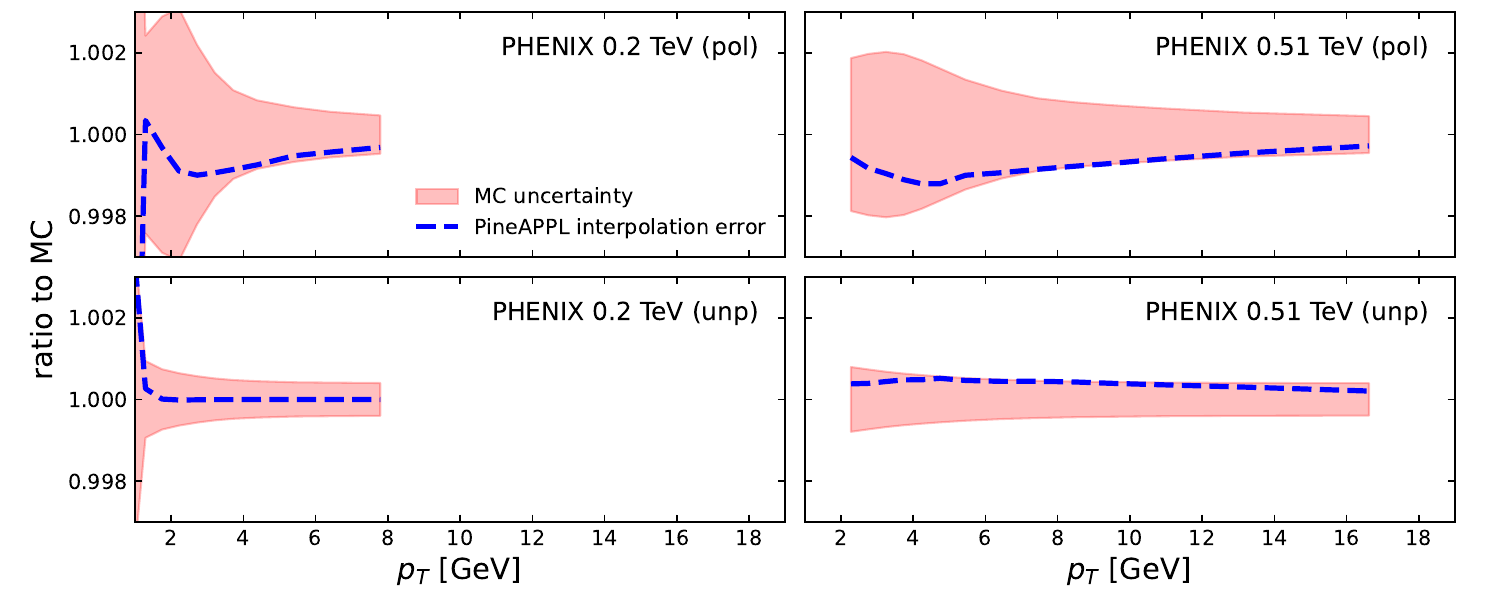}\\
  \caption{Same as Fig.~\ref{fig:unp_validation} for the polarised
    (upper panels) and unpolarised (lower panels) \pineappl\ grids generated to
    compute the double-spin asymmetries measured by the PHENIX
    experiment~\cite{PHENIX:2007kqm,PHENIX:2015fxo}, see
    Sect.~\ref{subsec:pheno-pp}.}
  \label{fig:pol_validation}
\end{figure}

Figures~\ref{fig:unp_validation}--\ref{fig:pol_validation} show the comparison
of the results obtained with the two aforementioned methods for each cross
section considered in Sect.~\ref{subsec:pheno-pp}, respectively in the
unpolarised and in the polarised cases. The uncertainty band corresponds to the
Monte Carlo integration error of the native code~\cite{Jager:2002xm}, whereas
the dashed line corresponds to the \pineappl\ interpolation relative error.
We make two crucial remarks. First, the Monte Carlo error is always very small
(of the order of permil or lower), especially if compared to the PDF, FF,
scale, and data uncertainties. This fact means that we have generated a
sufficient amount of integration events. Second, the \pineappl\ interpolation
error is always contained well within it. This fact means that there is not a
loss in accuracy due to the \pineappl\ representation and interpolation of
Monte Carlo weights on the grid. Both facts lead us to conclude that
computations are not spoiled by numerical inaccuracies at all.


\section{Installation and usage of \pineappl}
\label{sec:usage-pineappl}

In this appendix, we briefly describe the installation and usage of \pineappl,
which is available at
\begin{center}
  \faGithub~\href{https://github.com/NNPDF/pineappl}{https://github.com/NNPDF/pineappl}.
\end{center}
At this link, an interested user can find an exhaustive documentation about
installation instructions, tutorials, and code examples. These include
the \pineappl\ CLI and APIs, with explicit pieces of software in Python, C, C++,
and Fortran. In short, \pineappl\ can be installed via the Python Package Index
(PyPI), using any Python package installer. To install a given version of the
Python API and the \pineappl\ CLI one can use
\texttt{pip}:
\begin{minted}[fontsize=\small, linenos=false, breaklines, frame=lines]{text}
  pip install pineappl==x.y.z pineappl-cli==x.y.z
\end{minted}
where {\tt x.y.z} identifies the desired \pineappl\ version. The C/C++ API can be
installed using a pre-compiled binary:
\begin{minted}[fontsize=\small, linenos=false, breaklines, frame=lines]{text}
  curl --proto '=https' --tlsv1.2 -sSf https://nnpdf.github.io/pineappl/install-capi.sh | sh -s -- --version x.y.z
\end{minted}
The installation path can be specified with the \texttt{--prefix} flag. Once the
C/C++ API is installed, the Fortran API is automatically available through the
file \url{https://github.com/NNPDF/pineappl/blob/master/examples/fortran/pineappl.f90}. To use it, simply put the file in the working directory and
compile it with Fortran programs.

\pineappl\ is written in Rust, however, as mentioned, it has a C/C++ (with a
Fortran wrapper) API and a Python API. \pineappl\ also comes with a powerful CLI, which
we illustrate below. You can access all the API functionalities using the helper
as follows
\begin{minted}[fontsize=\small, linenos=false, breaklines, frame=lines]{bash}
  pineappl --help
\end{minted}
which prints all the available commands and options
\begin{minted}[fontsize=\small, linenos=false, breaklines, frame=lines]{text}
Read, write, and query PineAPPL grids

Usage: pineappl [OPTIONS] <COMMAND>

Commands:
  analyze   Perform various analyses with grids
  channels  Shows the contribution for each partonic channel
  convolve  Convolutes a PineAPPL grid with a PDF set
  diff      Compares the numerical content of two grids with each other
  evolve    Evolve a grid with an evolution kernel operator to an FK table
  export    Converts PineAPPL grids to APPLgrid files
  help      Display a manpage for selected subcommands
  import    Converts APPLgrid/fastNLO/FastKernel files to PineAPPL grids
  merge     Merges one or more PineAPPL grids together
  orders    Shows the predictions for all bin for each order separately
  plot      Creates a matplotlib script plotting the contents of the grid
  pull      Calculates the pull between two different PDF sets
  read      Read out information of a grid
  subgrids  Print information about the internal subgrid types
  uncert    Calculate scale and convolution function uncertainties
  write     Write a grid modified by various operations

Options:
      --lhapdf-banner          Allow LHAPDF to print banners
      --force-positive         Forces negative PDF values to zero
      --allow-extrapolation    Allow extrapolation of PDFs outside their region of validity
      --use-alphas-from <IDX>  Choose the PDF/FF set for the strong coupling [default: 0]
  -h, --help                   Print help
  -V, --version                Print version
\end{minted}
The interested user can test these commands with any of the \pineappl\
grids corresponding to the data sets discussed in Sect.~\ref{sec:pheno}. These
can be downloaded from
\begin{center}
  \faGlobe~\href{https://data.nnpdf.science/pineappl/pineapplv1/grids/}{https://data.nnpdf.science/pineappl/pineapplv1/grids/}.
\end{center}

\bibliographystyle{JHEP}
\bibliography{pineapplv1}

\end{document}